\documentclass[prl,aps,twocolumn,showpacs,floatfix,superscriptaddress]{revtex4}

\usepackage{color}
\definecolor{blue}{rgb}{0,0,1}
\usepackage{graphicx}
 \newcommand{\LATER}[1]{}
 \newcommand{\SAVE}[1]{}
 \newcommand{\QOK}[1]{}
 \newcommand{\QMM}[1]{}
 \newcommand{\MEMO}[1]{}
 \newcommand{\MMNOTE}[1]{}

\def\secc#1{{\it #1} ---}
\def\secchide#1{}
\def\la{\langle}
\def\ra{\rangle}

\begin{document}
\title{Caged clusters in Al$_{11}$Ir$_4$:
structural transition  and insulating phase}

\author{Marek Mihalkovi\v{c}}
\affiliation{
Institute of Physics, Slovak Academy of Sciences, 84511 Bratislava, Slovakia.}
\author{C. L. Henley}
\affiliation{Laboratory of Atomic and Solid State Physics, Cornell University,
Ithaca, NY, 14853-2501}

\begin{abstract}
Using pair potentials fitted to an ab--initio database,
combined with replica--exchange simulated annealing, 
we show the complex, quasicrystal-related Al$_{11}$Ir$_4$ compound 
contains a new version of the ``pseudo-Mackay'' icosahedral cluster, 
with non-icosahedral inner Al$_{10}$Ir and/or Al$_9$Ir clusters
that exist in various orientations and account for 
partial occupancies in the reported structure.
Two different compositions show first-order transitions to
orientationally ordered phases doubling the (cubic) unit cell,
which are respectively metallic and insulating.
\end{abstract}

\pacs{02.70.Ns,61.50.Lt,63.20.Dj,64.70.Kb,61.44.Br}

\maketitle

Al$_{11}$Ir$_4$ is a phase in one of the complex metallic alloy
systems that resist ordinary approaches to determining crystallographic 
structure or phase diagrams, because of two impediments to
equilibration at low temperatures:
(1) there are numerous competing complex phases of similar composition 
(2) the intermediate-temperature
structures have inherent entropy associated with 
{\it block rearrangements} of tiles or clusters.  

Despite the measurement of 457 independent reflections, the
experimentally refined structure~\cite{grin} of Al$_{11}$Ir$_4$ has
twice as many sites listed as atoms, many with a fitted 
occupancy of less than 0.5.
\SAVE{This means we do not know the space group of the actually
realized structure, and can not even trust its assignment to
a structure type.}
Until occupancy correlations are known, such structure knowledge
is inadequate for computing electronic properties or total energies.


Instead, in this letter we {\it predict} the structure, aided
by its close relationship to Al-transition metal (Al-TM) quasicrystals.
These are described as networks of identically oriented icosahedral clusters
linked along certain symmetry directions~\cite{elser-henley},
e.g. the ``Mackay icosahedron'' \cite{elser-henley}
consisting of an empty center plus two concentric shells
of full icosahedral symmetry, Al$_{12}$ + Al$_{30}$Mn$_{12}$.
\SAVE{These are packed on a bcc lattice (lattice constant 12.68~\AA) 
to form the crystal structure 
of $\alpha$-AlMnSi, or can be packed in a more general
way to form the quasicrystal $i$-AlMnSi.}
It is generally accepted that the stable quasicrystals 
$i$-AlCuFe and $i$-AlPdMn contain a modified cluster
called the ``pseudo Mackay icosahedron'' (pMI)
in which the inner Al$_{12}$ icosahedron is replaced
by a cluster with fewer atoms and reduced symmetry~\cite{katz-gratias},
the details of which are still unclear.
We note that the Ca--Cd quasicrystals, another major class
alongside the Al--TM quasicrystals,  and complex alloys related to them,
also contain icosahedral clusters with an inner shell 
of lower symmetry~\cite{gomez} -- a tetrahedron, in that case. 
Thus, such inner rotatable clusters may be a common feature 
of well-ordered quasicrystal systems, and they present tractable 
examples of the above-mentioned block-rearrangement degrees of freedom.

In the rest of this paper, we uncover the correct low-$T$ structure
of Al$_{11}$Ir$_4$ using energy minimization, in several stages.
We used both ab-initio codes (VASP package~\cite{vasp}) and empirically 
fitted pair potentials.  First we determined the possible atom contents 
of one cluster to be Al$_9$Ir or Al$_{10}$Ir. Next, for each cluster,
we find its atom arrangement and preferred orientation
within the cage of neighboring atoms.  The partially occupied sites in 
the refined structure~\cite{grin} are explained with an equal 
mixture of the two kinds of cluster.  We also find the
collective ordering of orientations and identify the ordering transition.
\SAVE{(That is, its temperature and nature (first-order or continuous?).}
The equal-mixture phase is metallic, as seen in experiment~\cite{tamura},
but the composition with Al$_{10}$Ir clusters is predicted to be insulating.

\secchide{Fitted empirical oscillating pair potentials}

As a prerequisite to exploring alternative structures and running 
extensive molecular dynamics (MD) simulations,
we generated ``empirical oscillating pair potentials'' (EOPP)
valid for Al--Ir in this composition range, 
using the method of Ref.~\cite{MM-ppfit} to
fit a database of ab--initio forces, calculated with VASP~\cite{vasp}. 
The database contained snapshots from ab-initio molecular dynamics
simulations at high temperatures, as well as 
$T=0$ structures of various stable and unstable Al--Ir structures,
in particular the Al$_{21}$Pd$_8$ and $B2$ structures. (The functional
form and fitted coefficients of the potentials are
provided in the supplementary material~\cite{supplementary}.)
\SAVE{(MM 11/25/11)
MD samples : Al11Ir4 2x2x2 supercell at 1700K, pure Al 300K, pure Al 600K.
Quenches from 1700K, Al11Ir4 2x2x2 supercell.
Relaxed T=0K samples: Al9Co2.mP22, Al3Ir.hP8, AlIr.cP2, Al5Co2.hP28, Al45Ir13.oP236.}
\SAVE{Reference for Al$_{28}$Ir$_8$: 
S. Katrych, V. Gramlich, and W. Steurer,
J. Alloys Compds. 407, 132-140 (2006),
"Trigonal Ir$_9$Al$_{28}$, a new structure type and approximant
to decagonal quasicrystals"}

\secc{Structural optimization of individual cluster in Al$_{11}$Ir$_4$}
\SAVE{The reported crystal structure of Al$_{11}$Ir$_4$ is 
a modification of the ``AuZn$_3$'' structure, 
with Au$\rightarrow$Ir, Zn$\rightarrow$Al, and some partial occupancies.
The AuZn$_3$ structure, has  Pearson symbol $cP32$),
It is a a bcc packing of Ir icosahedra, distorted
so as to share faces, and having icosahedral 2-fold axes aligned with the 
cubic 4-fold axes (there are two equivalent ways,
one used by the body-center cluster and one by the vertex cluster, 
thus the space group is simple cubic 
$Pm\bar{3}n$ with a fourfold screw axis.
Inside each is a smaller, less distorted AuZn$_{12}$ icosahedron 
aligned with the cubic 4-fold axes in the other way
than its Au$_{12}$ cage.}
The reported crystal structure of Al$_{11}$Ir$_4$ 
is a simple cubic arrangement of pMI clusters, each 
having an inner shell of $\sim 40$ partially
occupied sites with \SAVE{(reported)} combined occupancy 10.
Surrounding this is the second shell, an Ir$_{12}$ icosahedron, 
plus 30 Al sites slightly outside that icosahedron's mid-edges,
of which the six sites along cubic 2-fold axes are
only partially occupied.
\SAVE{(fully occupied along 24 non-cubic icosahedral 2-fold axes)}
\SAVE{Thus each body centers has an MI cluster with 
very distorted outer shells, that are shared with inner shells
of the adjacent pMI.}
Between the pMI clusters, at each body center,
is an Ir atom, surrounded by an icosahedron of 12 Al atoms,
which are shared with the pMI in the role of second-shell Al.
The key fact to determine by energy optimization 
is the atom configuration in the uncertain,
inner-shell sites.
\SAVE{The interaction between the orientations 
of neighboring clusters is a small perturbation.}


\begin{figure}
\vskip 1.0cm   
\includegraphics[width=3.2in]{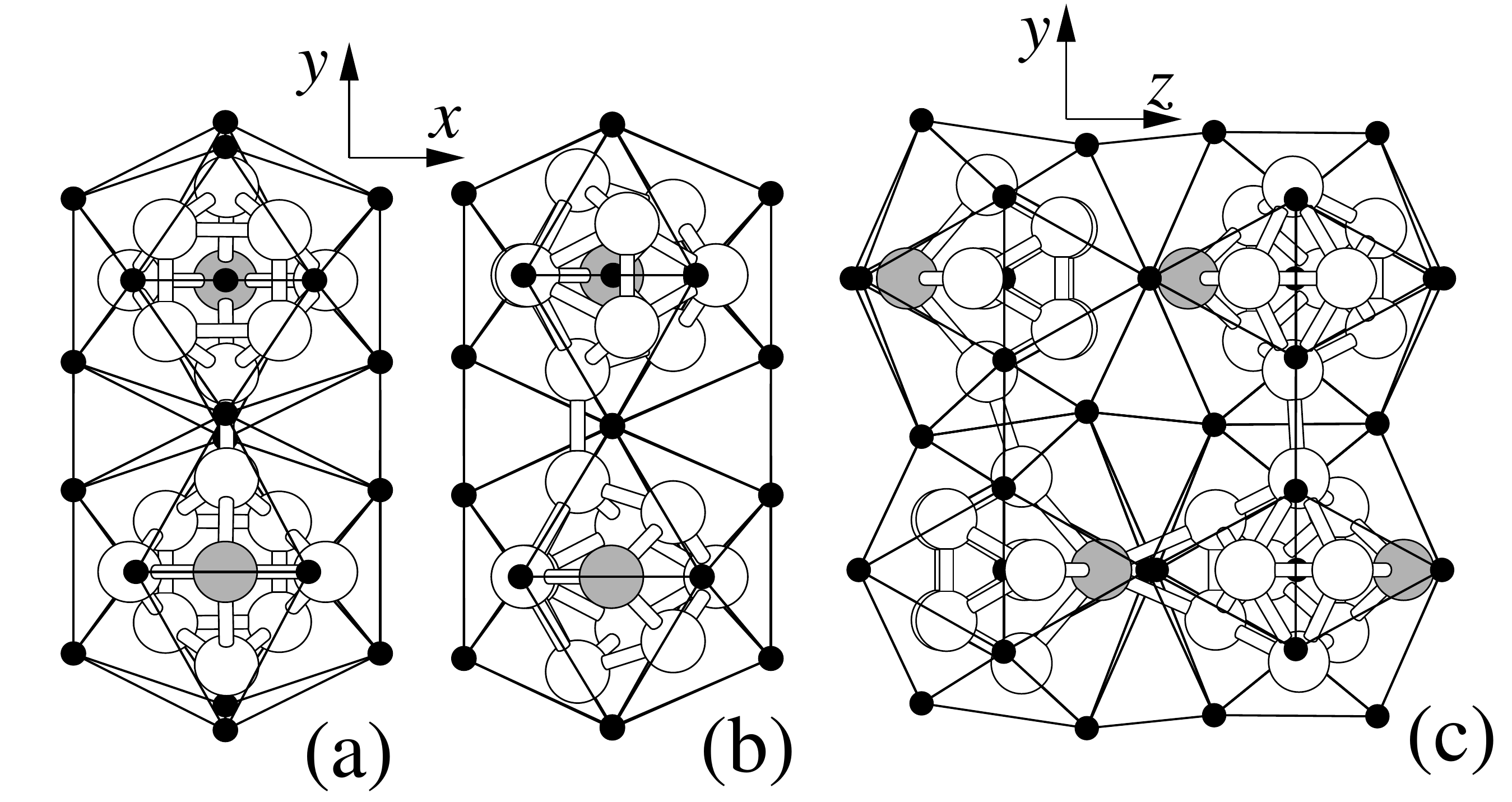}
\caption{
Inner clusters in  Al$_{11}$Ir$_4$ structures,
shown as they order in the low-$T$ ``9.5 phase''.
The surrounding icosahedron of black atoms is the Ir$_{12}$ ``cage'';
another Ir atom is at center. The ``apex'' Al atoms are shown gray.
\SAVE{CLH wonders if the caging icosahedron should be
modified to show the mid-edge Al atoms by tiny open circles,
wherever these are not capping atoms of the inner cluster;
thus many edges of it would become broken into two
segments.  MM: no quick way to show only those Al-Ir bonds
along icosahedron edges. CLH: ? pretend those are a
different flavor of Al? It would work for the 24 of the 30 edges.
But not for the cubic 2-fold atoms which are a wrapped images
of the inner-Al's, if they appear there at all. I just thought it's
too complicated.}
(a). Al$_{9}$Ir clusters in $z=0$ plane of the ``9.5 phase'' with
their 4-fold axes oriented up and down along viewing $z$
axis,
(b). Al$_{10}$Ir clusters in $z=0.5$ plane, with their 3-fold axis almost parallel
to the viewing $z$ axis 
\SAVE{(Notice how the foreground side of the lower cluster in (b) 
is a half icosahedron)}.
(c). Orientation relation of two Al$_{9}$Ir clusters (left)
with two Al$_{10}$Ir clusters (right), viewed along $x$ direction.
The layers of identical-content clusters are stacked normal
to the page in (a) and (b), or left to right in (c), in a 
checkerboard pattern. In the
10 phase, 10-cluster arrangement in (b) is repeated in 2$\times$1$\times$1
supercell.
\SAVE{In (a), there is a twofold 
axis normal to the page midway between the left two clusters,
and another midway between the two right clusters.}
}
\label{fig:clusters}
\end{figure}


The first question about the inner Al cluster
is how many atoms it (optimally) contains. We addressed
this directly by constructing initial configurations
of the Al$_{11}$Ir$_4$ unit cell with a chosen number 
$m$ of inner Al atoms (assuming the periodicity of only one unit cell).
The configuration was optimized, first by annealing
with molecular dynamics (MD)
using the fitted EOPP potentials at fixed volume; 
this was followed by ab-initio relaxation using VASP~\cite{vasp},
optimizing all structural parameters.~\cite{FN-supercells}.
\LATER{CLH has second thoughts about hidden line:}
\SAVE{
The atoms within the inner shell rearrange rapidly
under MD, whereas the surrounding cage atoms merely vibrate, 
so the end state approximates the optimum 
constrained to the chosen content $m$ Al 
atoms in the cell-corner clusters and $12$ Al atoms in the body-center clusters.}
The first stage was annealing in one unit cell
(see Table~\ref{tab:cluster-energies}), to identify gross energy
differences: this excluded the $m=8$, $m=11$ and $m=12$ variants,
\SAVE{(The $m=12$ case is identical to the basic bcc AuZn$_3$ structure.)}
leaving only the $m=9$ and $m=10$ ones
which we ``9-cluster'' and ``10-cluster'' from here on.
In a second stage, we annealed in a supercell to accomodate
possible alternations of the orientations, and found two
nearly stable structures: the $m=10$ filling (``10-phase'') 
and an equal mixture of $m=9$ and $m=10$ fillings (``9.5-phase'').
From here on, in studies of the whole structure,  
we limit ourselves to these two fillings.
\SAVE{Notes 12/3/2011 from MM:
All these energies are NEW, because the tie-line has been changed, 
the main stable competitor is Al21Pd8.
(i) stage 1: annealing within UNIT CELL to find gross differences.
The 9-atom model is unstable by 25 meV/atom, 11-atom by 77 meV/atom.
12-atom by 132 meV/atom (AuZn3.cP32 type structure).
The 10-atom model was +10 meV/atom. 
(ii) stage 2. 2x2x2 supercell annealing (9-atom and 9.5 atom models), 
and 4x4x4 supercell annealing for 10-atom model. Stability (new tie-line Al45Ir13-Al21Pd8):
10-atom : 2x1x1 supercell, 60 atoms, ``Al11Ir4.oP60'', +3.8 meV/atom
9.5 atom: 2x2x2 size cell, centred ortho. 236 atoms, ``Al11Ir4.oA236'', +9.4 meV/atom
9-atom : 2x2x2 size cell, ``Al11Ir4.oI232'' (232 atoms), +18.2 meV/atom 
This new tie-line  moved especially 9-atom model up in dE, because it has
composition very close to Al21Pd8. But it also destabilized 10-atom model,
which is now +3.8 meV/atom! This should not bother us in this story, our
structure will apparently have large entropic advantage.
}

\begin{table}
\begin{tabular}{l|cccccc}
inner cluster Al     & 8 &    9 & 9.5  &  10  &  11 &   12    \\
\hline
$1\times 1 \times 1$ &+45& +33  & --   & +10  & +77 & +132  \\
  supercell          &--&+18   & +9   &  +4  & --  &  --   \\
  $a_{cub}$ [\AA]     &7.64&7.68  & 7.70 & 7.73 & 7.78 & 7.82 \cr
\hline
\end{tabular}
\caption{
\label{tab:cluster-energies}
Relaxed energies (in meV/atom) as a function of Al per pMI inner cluster.
Supercell used was $2\times 2 \times 2$, except 
$4\times 4 \times 4$ for the 10 Al case. Last row is mean lattice parameter,
per fundamental cell.
\SAVE{MM 12/14/11: changed original entry of +25 meV/atom for unit cell 
9-phase to +33 meV/atom, because
this state actually transformed into 10-cluster, borrowing one
atom from BCC icosahedron; redid annealing at lower temperature and
indeed the 9-cluster survived, energy +33 meV/atom. This is however not the
``cubic'' 9-cluster we describe, it is rather 10-cluster without the 10th
capping/apex atom, as in $\ksi$-AlMnPd. So this is metastable 
configuration.}
\SAVE{The lattice parameters relaxed by VASP should NOT be directly compared
to the experimental value, $a$=7.674\AA, because there is a systematic error.
Instead, we have to ``calibrate'' expe. vs. VASP by using an exactly known
reference strructure. Such convenient structure here is Al45Ir13 (oP236), which is
computed to be stable, is indeed observed stable experimentally, and is Al-rich.
For this structure, VASP cell parameters are about 1.006 multiple of the experimental
parameters. So the expe. value should be multiplied by this correcting factor, in order
to compare with the last row of the table: $a_{ref}$=1.006$\times$7.674=7.720\AA.
And this value is exactly between our 9.5 and 10 phase cases!
}
\SAVE{
The actual emergent structure was $2\times 1 \times 1$ for the
10 Al case.}
}
\end{table}

The 9-cluster is almost always a square antiprism, with one square
enlarged and capped by an additional Al atom
[see Fig.\ref{fig:clusters}(a)].
\SAVE{(Equivalently, it could be considered a square pyramid with
capping atoms on the three triangular faces.)}  
In its optimum 
orientation (of multiplicity six), the fourfold axis is aligned
with one of the cubic $\la 100\ra$ axes, with the outer
square of atoms in mirror planes.
\SAVE{The five outer atoms of the 9-cluster 
are aligned [Fig.~\ref{fig:clusters}(c), pointing
toward cubic 2-fold directions, where they share
the role of second shell cage Al. 
Thus, the only icosahedron edge not having a cage Al near 
its mid-point is the one near the apex atom of Al$_9$Ir cluster.
}
The 10-cluster is practically always a sort of trigonal prism, 
with one triangular end face enlarged and capped by an atom,
and also capped on the three trapezoidal side faces.

\SAVE{
An alternate way to visualize the 10-cluster is as a 
reconstruction of a complete inner Al$_{12}$ icosahedron,
rotated 90$^\circ$ with respect to the Ir$_{12}$ cage
icosahedron
(as found in the AuZn$_3$ structure, of which Al$_{11}$Ir$_4$
is a modification).
First choose any (fivefold) vertex of the cage icosahedron, which 
is approximately aligned with a certain threefold axis of
the inner icosahedron: that becomes the threefold axis for the 10-cluster.
Replace the triangle of three Al around that axis
by a single Al, which becomes the apical capping atom; 
the three atoms next closest to that triangle move
up towards the apex, and the three atoms after that move out
radially a bit, to become the side-capping atoms of the trigonal prism.
\LATER{Drop this sentence if necessary (MM said OK)}
The effect is that the ``top'' hemisphere is approximately
half a cube (see Fig.~\ref{fig:clusters}(b), lower cluster), 
the ``bottom'' hemisphere (opposite to the 
chosen direction) remains a half-icosahedron (upper cluster in the
same Figure), and they
share a puckered ring of six atoms.
}

In its optimum orientation (of multiplicity 12),
the 10-cluster is tilted rigidly so as to bring its threefold 
axis closer to (but not quite parallel to) the nearby $\la 100\ra$ axis,
as seen in Figure~\ref{fig:clusters}(b).
That brings the apical capping Al and all three of the side-capping Al atoms
(which are farther out from the central Ir) close to cubic $\la 100\ra$
axes, and they merge with the Al atoms of the pMI's
second shell (cage) that sit in the same direction.
The pMI clusters in Fig.~\ref{fig:clusters} {\em overlap}
along the $[100]$ direction 
such that some outer 
shell atoms of one pMI also belong to the inner cluster 
of the neighboring pMI~\cite{FN-elser-pMI}.
\SAVE{(along the $[010]$ direction, they are linked by
a pair of capping atoms; along the $[001]$ direction,
the apex atom from one cluster togther with two atoms
from the other make a triangle.)}

At $T\cong 600K$,  just 100K above the orientational $T_c$ 
(see below), an inner Al$_{10}$Ir cluster spends 95\% of the time close to one 
of the twelve ideal orientations described above (as shown by quenching
a crystal using Al$_{10}$ filling in each cage.)
At higher temperatures, 
the threefold axis of the 10-cluster may instead align along 
a cubic $\la 111\ra$  axis, or a pseudo 2-fold axis (of the Ir$_{12}$ cage).
(We have not examined the high-$T$ orientations of the 9-cluster.)
\SAVE{The discrete additional orientations 
are presumably local minima of higher energy.
Sometimes, one inner-cluster
Al atom may wander into a neighboring pMI shell, 
so the respective clusters are Al$_9$Ir and Al$_{11}$Ir,
like a vacancy-interstitial pair; this might be
a mechanism for Al self-diffusion.
The body-center MI are not involved in this disorder;
the cell-corner pMI and body-center MI clusters
are distinct even at high $T$.}

\SAVE{The {\it body--centers} of the cubic lattice are still filled 
by IrAl$_{12}$ icosahedra [not shown in Fig.~\ref{fig:clusters}]}

\begin{figure}
\vskip 1.0cm   
\includegraphics[width=3.2in]{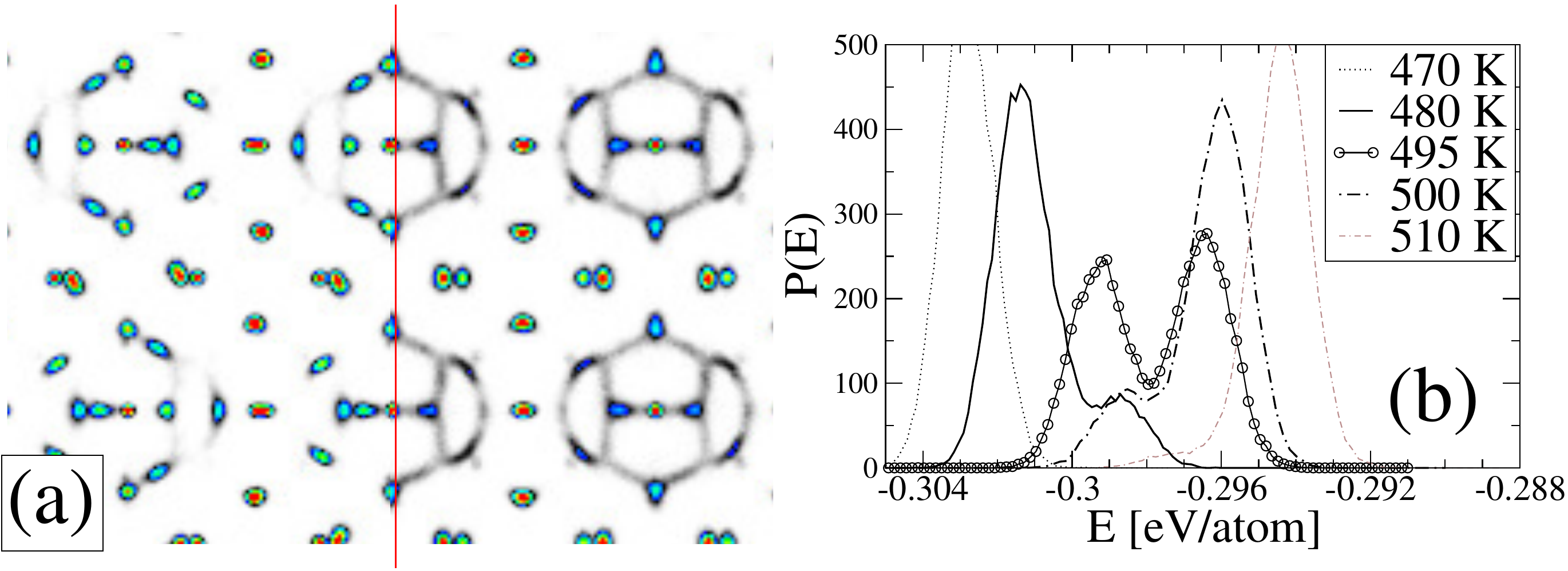}
\caption{Disordered and ordered states of inner clusters.
Phase transition in the ``10'' phase from replica--exchange molecular dynamics
simulation in $4\times 4\times 4$ supercell of the basic (30--atom) unit cell.
simulation results from $4\times 4\times 4$ 
supercell of the basic (30--atom) unit cell. 
(a) time--averaged occupancy distributions for a slice 
of thickness 6\AA~ taken through the $4\times 4\times 4$ supercell
of the ``10'' phase composition, below the ordering transition
at $T=$470 K (left) and above it at 510 K (right).
\QOK{CLH hid this line anyhow, seemed obvious enough.}
\SAVE{Time--averaging restores the ordered structure's cubic symmetry.}
(b) Probability distribution $P(E)$ for total energy (normalized per atom)
of the ``10'' composition, at five temperatures from 470K to 510K;
the double peak at 480 K $\leq T \leq$ 500K is diagnostic of
coexistence and a first-order phase transition.
\SAVE{(The variance of this distribution is proportional to the
heat capacity.  The two peaks are separated by about  3 meV/atom, 
the relative weight of which shifts with $T$.}
}
\label{fig:md}
\end{figure}

The Al partial occupancies refined in \cite{grin} are
in close agreement with our model, provided that 
close to half the clusters are 9-clusters and half are 10-clusters,
and each is oriented independently in one of the (respectively)
six or 12 optimum orientations.~\cite{FN-Grin-spacegroups}
(Fig.~\ref{fig:md}(a)
shows the time-averaged atom density from a high$-T$
simulation, containing lines of closely spaced alternative sites.
The refined Al(3) and Al(4) sites of \cite{grin} represent
this spread-out density, coming from four or five
model sites of both 9- and 10-clusters; the reported Al(2)
and Al(5) sites comes respectively from one site of the 
9-cluster and two sites of the 10-cluster in our model,
showing that both clusters must be present.  The r.m.s. 
discrepancy between predicted and reported occupancies of each
site is $\sim$12\% of the occupancy~\cite{supplementary}.


\secc{Orientational ordering} 
To address the statistical mechanics of systems with many
interacting clusters, we set up a replica--exchange~\cite{replica-exchange} 
simulation in a $4\times4\times4$ supercell, 
i.e. 1920 atoms (10-phase), annealing 16 independent samples, each at a
different temperature, spanning 310K to 460K; additional
single-$T$ simulations were done at higher temperatures.

The simulated alloy remains solid beyond 1700 K in 2x2x2 supercell (the experimental
melting point is above 1600 C for $x_{Ir}\sim0.27$ ~\cite{oka}).
The simulation lasted 4000 cycles, each consisting of 
1000 molecular dynamics (MD) steps with time increment $\Delta t$=5 fs,
for a total simulation time 20 ns.
\SAVE{Each cycle is an opportunity to exchange the temperatures of two samples.}
The  pair-potential
interaction radius was cut off at $r_c$=10\AA.~\cite{FN-rc}.
The initial state for the MD was our best single-cell model, 
repeated in all $4^3$ cells.

The {\it low--temperature} optimal structures of the 9.5-phase and 10-phase
were determined by rapidly quenching configurations from
the lowest replica ($T=310$K) to a $T=0$ energy minimum,
\SAVE{(In 50 iteration steps)}
and selecting those with the lowest relaxed energies. 
In 9.5-phase, the structure was a stack along (say)
the $z$ direction of layers which had a $\sqrt{2} \times \sqrt{2}$ 
checkerboard pattern of inner clusters, with their main (fourfold or
threefold) axes alternating
between $+z$ and $-z$ directions; the layers are stacked
such that the overlaying clusters have the same orientation
of main axes.  (To fully specify their orientations,
note the 10-clusters of $+z$ and $-z$ alignment 
are always related by a mirror reflection in the $xy$ plane.)
The relation between adjoining clusters is shown in Figure~\ref{fig:clusters}.
In the 10-phase, the 10-clusters pair shown in 
Fig.~\ref{fig:clusters}(b) is repeated in the supercell doubled along $y$ direction.
The crystal structures are summarized in Table~\ref{tab:crystal}; coordinates
are available in Ref.~\cite{supplementary}.

\SAVE{
\begin{figure}
\vskip 1.0cm   
\includegraphics[width=3.2in]{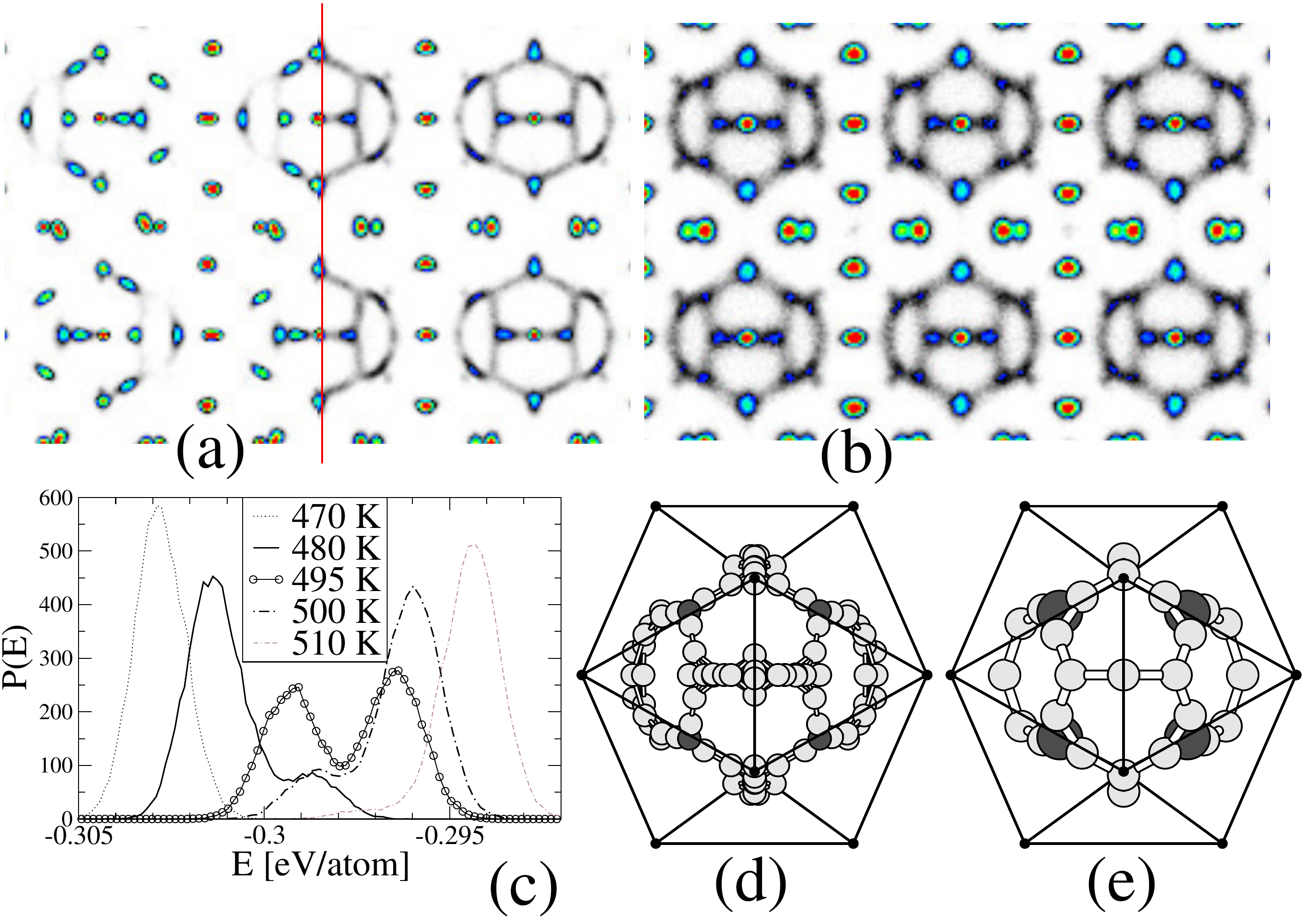}
\caption{Average state of inner cluster in ``9.5'' phase
(a) Same plot as Fig.~\ref{fig:md}(a), but
for the ``9.5'' phase composition at $T=$1173 K;
compared to Fig.~\ref{fig:md}(a),
we see a strong density along
the $\la 111\ra$ directions.
\SAVE{(simulation supercell is also $4\times 4\times 4$}
(b,c) Partially occupied inner Al sites in each pMI cluster
in the low-temperature 9.5-phase structure;
sites along $\la 111 \ra$ directions are shown dark.
(b) Union of sites in our model structure;
\SAVE{six from the 9-cluster and twelve from the 10-cluster)}
(c) sites in diffraction refinement of Ref.~\cite{grin}.
\SAVE{Here (c) is an average over domains of the low T
ordered structure; not the same state as (a), though it 
looks similar.  MM notes, at 1170K 
there are other shapes of 9-cluster (eg like the
one in ksi-AlMnPd), there are
pairs of 10-cluster and 8-cluster, or even 9-clusters
interact with BCC Al12Ir to borrow one atom. That's 
what's puzzling about Tamura's Pa3bar state.}
}
\label{fig:inner-sites}
\end{figure}
}

\begin{table}

\begin{tabular}{lllcccrr}
 structure &  Pearson  & Space  &  \multicolumn{3}{c}{parameters (\AA)} & Al & Ir \\
           &  symbol   &  group &  $a$    &  $b$  &   $c$  &   &  \\
\hline
Ref.~\onlinecite{grin}  & cP60   & $P23$    &  7.67 & \multicolumn{2}{c}{(cubic)} & 22  & 8 \\
high-$T$ 9.5  & cP178   & $Pm\bar{3}$    &  7.73 & \multicolumn{2}{c}{(cubic)} & 21.5  & 8 \\
9.5-phase     & oA236   & $Abm2$         & 15.48 & 15.41 & 15.32           &  172  & 64 \\
10-phase      &  oP60   & $Pma2$         & 15.49 & 7.74  & 7.70            &  44   & 16 \\
\hline
\end{tabular}
\caption{ 
\label{tab:crystal}
Crystal data for Al$_{11}$Ir$_4$ structures,
from diffraction refinement~\cite{grin}, the disordered
high-$T$ 9.5-state and the ordered states of the 9.5-phase
and 10-phase.  
\QMM{I don't see how we can avoid telling what the
numbers in the table columns mean, since the column
headings don't say}
Columns include lattice constants and Al or Ir atoms per cell.
For ``high-$T$'' 9.5, data in ``(...)'' are from Ref.~\onlinecite{grin}.
\QMM{``Row ``high-$T$'' 9.5 refers to average over
9-cluster and 10-cluster orientations.''  I have no idea what this
means!  Are you saying it is actually the low-$T$ 9.5 phase, with
this averaging?  But my main concern is the composition of
the high-T 9.5 phase: shouldn't it be the same
as the low-T 9.5 phase?  I get 21.5 Al and 8 Ir.}
\SAVE{Why CLH put in the ``number of sites''.
In the first place, I had no other place to look for the number of
orbits.  But mainly, I just intended for someone who was trying
to replicate our work to identify what structure we were talking
about, and to say ``yes, this must be the same as the one they
had in mind.''}
\SAVE{The site list corresponding to the high-$T$ 9.5-phase 
and Ref.~\onlinecite{grin} are depicted in Fig.~\ref{fig:inner-sites}(b) 
and (c), respectively}
\SAVE{Grin's refinement has 0.5-occupied Ir icosahedron sites.
This is Ir 0.5 and vacancy 0.5.  
MM: It is unusual to have fractionally occupied TM site in these alloys,
especially Ir/vacancy (as in Grin's structure) calls for explanation.
But it does not matter here: that Ir site only
has double multiplicity (and is 0.5 occupied) but we wouldn't 
need to tell that in a table.}
\SAVE{(MM believes it is due to Ir icosahedron distortion with 9-clusters.)}
\SAVE{9.5-phase : a=15.48, b=15.41, c=15.32 (as relaxed by VASP).}
}
\end{table}

\SAVE{Inner 10-clusters in adjacent clusters,
related by a cubic $\la 100 \ra$ vector,
have the following relation.
Remember that pMI clusters {\em overlap}
along the $\la 100\ra$ directions. 
Towards the $\la 100\ra$ directions that do {\it not} have a capping
atom, the 10-cluster instead shows a pair of Al atoms, aligned
roughly in an $\la 011\ra$ direction.
It is favorable for this pair to face a capping Al on the adjacent 10-cluster, 
so as to form a triangle of Al atoms.  Such a relationship occurs 
along the $[001]$ axis and $[010]$ axes as shown in 
Figure~\ref{fig:clusters}.}

\SAVE{
To continue: in a pair of 10-clusters related by the $[001]$ axes,
{\it each} cluster is showing a pair of Al, so it is sterically
favorable for the clusters to be oppositely aligned
such the 10-clusters are linked by an Al$_4$ tetrahedron.
(I'M NO LONGER SURE THE TETRAHEDRON IS AN ACCURATE 
PICTURE OF THE CONFIGURATION.)
This explains why the clusters related by $x$ linkages
have alternating orientations,
as shown in Figure~\ref{fig:clusters}(c).}

\LATER{check if next paragraph duplicates the same info elsewhere.}
The ordered arrangement of the 9.5 phase contains
alternating (square) layers of 9-clusters and 10-clusters.
Each layer is made by repeating Fig.~\ref{fig:clusters}(a) or (b)
in both directions: thus, within either kind of layer, 
the apex directions alternate up and down,
These layers are stacked without shifts so that chains of 
identically aligned clusters run perpendicular to the
plane of the layers.

\secchide{The phase transition}

We pinpointed the ordering temperature of the ``10'' phase in three ways.  
Firstly, we can
examine the atom density distribution, averaged over moderate times
(Fig.~\ref{fig:md}(a); for $T>T_c$ this has the full symmetry
of the unit cell, but for $T<T_c$ this shows (despite some fluctuations)
a clear symmetry breaking (to the cell-doubled structure just described).
A second evidence for a phase transition is available within
the replica-exchange simulation: a larger energy spacing
$\Delta \bar{E}_i = E(T_i)-E(T_{i-1})$ between the replicas
at consecutive temperatures, suggesting a latent heat.
The third and most convincing signature 
is that, for a finite system, the ensemble at temperatures close to T$_c$ is
a mixture of the two phases, weighted according to the difference in their
free energies.  This is evident in Fig.\ref{fig:md}(b), from which
we can read off $T_c  \lesssim 495K $.
The peak separation in Fig.\ref{fig:md}(b) shows that the ordered and disordered
states in the 10-phase differ by
(a latent heat of) 3 meV per atom, requiring an entropy
difference at $T_c$ of (30)(3 meV)/$T_c \approx \ln(9)$
per 30-atom cell.  That is, we have effectively $N_{cs}\approx9$
states per cluster, comparable with $N_{cs}=12$ ideal orientations.

\SAVE{We also tentatively concluded there are no phase transitions
(in the 10-phase)
at higher $T$ by performing MD simulation at temperatures
600-2000(??) K, and examining the symmetry of the
time--averaged densities visible in Fig.\ref{fig:md}(a)  and (b).
With increasing $T$, the streaks in Fig.\ref{fig:md}(b) get stronger, 
but they do not change in shape.}

The 9.5 phase has two kinds of order -- the 9Al/10Al alternation
in cluster content and the orientations -- which  might appear in separate transitions.
The first ordering requires vastly longer equilibration times 
(for inter-cluster Al diffusion) and we were unable to 
identify any sharp transitions.
\SAVE{CLH notes, the 9.5 phase at high $T$ has effective
$N_{cs}=2 \sqrt{(6)(12)}\approx 17$: a factor of two for deciding
which clusters get the 9.5 or 10 filling, and six orientations
for each 9 cluster.}
\SAVE{
In principle, we might see multiple ordering transitions.
For example, conceivably in the 10 phase the main axes might
align along $\pm z$ while the secondary direction (shown by which 
way the triangle is oriented when projected on $xy$) is still
disordered.  Or, in the 9.5 phase, the main axes might
align along $\pm z$ while the 9 and 10 clusters are randomly
distributed across layers.  (The segregation of 9 and 10 clusters
into layers seems to be driven by orientational interactions,
so it seems less plausible to have the converse situation of
segregational order with orientational disorder.)
MM also notes the possibility of a transition between the 
9.5 and the 10 phase.}

\secc{Phase stability in Al--Ir system}
\LATER{We agreed to condense the next paragraphs,
but CLH did not see how.  
  or change the emphasis, since up to now the paper is all about
  Al$_{11}$Ir$_4$ and finding the correct low-T state; suddenly we are
  talking about all these phases on an equal footing, e.g. how we
  treated partial occupancies.}
\SAVE{(MM calls that ``hP24-1''.)}
\SAVE{The Al--Rh system also has the Al11Ir4 structure; in
that system the Al5Co2 structure IS stable.}
We now turn to the $T=0$ phase stability of Al$_{11}$Ir$_4$ and 
other Al-Ir compounds.
The currently accepted Al--Ir phase diagram~\cite{oka} 
with $0< x_{\rm Ir} \leq 0.5$ shows
six compounds as stable:
\SAVE{(Ref.~\cite{oka} does not use the name
Al$_{11}$Ir$_4$ but instead calls it
 but instead calls it ``Al$_{2.7}$Ir''.)}
Al$_9$Ir$_2$ (in the Al$_9$Co$_2$ mP22 structure); AlIr; 
and around $x_{\rm Ir} \approx 1/4$, there is Al$_3$Ir
plus the complex phases 
Al$_{11}$Ir$_4$, Al$_{28}$Ir$_9$~\cite{katrych2006} and orthorhombic Al$_{45}$Ir$_{13}$~\cite{al45ir13}.
We additionally tried other complex phases structures not reported in Al--Ir, among them :
\SAVE{ We also tried Al$_5$Ir$_2$ in the Al$_5$Co$_2$ structure
(like Al$_5$Rh$_2$ ~\cite{Al5Rh2}.}
(i) Al$_{21}$Ir$_8$ in the Al$_{21}$Pd$_8$ structure,~\cite{Al21Ir8}
which is a packing of sixteen of our 10-clusters per unit cell,
{\it without} any icosahedral cages (and sharing a few Al atoms).
(ii) Al$_{41}$Ir$_{23}$ in the 
Al$_{41}$Cu$_8$Ir$_{15}$ structure~\cite{AlIrCu},
equivalent to 
Al$_{68}$Pd$_{20}$Ru$_{12}$~\cite{mahne-steurer} 
and AlCuRuSi~\cite{sugiyama-AlCuRuSi};
here the even-vertex clusters alternate between 
an Al$_{10}$Ir trigonal cluster and an AlCu$_8$Al$_6$ cluster,
identical to atomic arrangement in $B2$ cubic structure.
\SAVE{MM 12/2011: the ``stoichiometric'' composition is Al41(Cu,Pd)8Ir15.
MM tried this same
structure substituting Cu$\rightarrow$Pd like Tamura et al reported, and
it WAS STABLE in the Al-Ir-Pd ternary, composition Al41Ir15Pd8.
NExt, substituting Cu$\rightarrow$Ir also works giving Al41Ir23, which is
only +6 meV/atom unstable.}
\SAVE{The trigonal clusters occur in a $2\times 2\times 2$ supercell in
all (eight?) possible orientations, giving a Pa$\bar{3}$ structure
as reported by Tamura.}

Of these, the available structures of Al$_{11}$Ir$_4$ and Al$_{28}$Ir$_9$ 
include many fractional sites, so in total energy calculations 
we had to try versions of these structures with various ways of
realizing the occupations.  
\SAVE{Orthorhombic Al$_{45}$Ir$_{13}$ (Pearson symbol oP236), is complex, 
but has only, at one place, a single PAIR of sterically exclusive sites.}
The relaxed total energies of all these compounds
(and the pure elements) were computed using VASP~\cite{vasp}.
\SAVE{(The convex hull of the energy/composition plot was constructed.
using the package ``Qhull''~\cite{qhull}.)}
We found AlIr and Al$_9$Ir$_2$ to be stable, as expected;
the only stable phases around $x_{\rm Ir}\approx 1/4$
were Al$_{45}$Ir$_{13}$ and (surprisingly) Al$_{21}$Ir$_8$.
Relative to a corrected tie-line including Al$_{21}$Ir$_8$,
we found Al$_3$Ir to be unstable by 25 meV/atom;
this is reduced to only $\sim 8$ meV/atom in a variant
(Al$_{17}$Ir$_{6}$) with a tripled cell in which one Al is removed.
Thus, in contradiction to Ref.~\onlinecite{oka}, 
we believe Al$_3$Ir is a high-temperature phase only.
The phases  
Al$_{41}$Ir$_{23}$,
Al$_{28}$Ir$_9$, 
and Al$_{11}$Ir$_4$ in
the 9.5-phase were unstable by rather small amounts of 6, 7, and 
9 meV/atom, respectively; 
\SAVE{(2.9 meV/atom for Al$_5$Ir$_2$, 3.8 for Al11Ir4-9.5)}
in the latter two, the site disorder contributes
entropy which may explain their stability  at higher temperatures.
\SAVE{Al28Ir9 is the isostructural
phase with the new AlPdRe phase by Sugiyama et al. 
As noted above, due to site occupancy disorder, this phase
has an obvious entropic potential for stabilization.
On the other hand, Al$_{21}$Pd$_8$ is a fully occupied structure..}
The heat of formation for Al$_{11}$Ir$_4$ is large: $-0.738$ eV/atom;
for Al$_{21}$Ir$_8$ it is $-0.760$ eV/atom (see Supplementary material
for detailed information on formation energies and stability data for
all discussed phases).
\SAVE{We note that Al$_5$Ir$_2$ is predicted to have
a large heat of formation, -0.765 eV/atom; 
Nevertheless this (Al5Co2) structure is unstable because it's suppressed 
by tie-line with AlIr B2 phase with -0.955eV heat of formation.
It is unstable yb 3 meV/atom.}
\MMNOTE{MM was looking for an alternate place to reference
Tamura's paper.}
\SAVE{12/2011: at one point recently (but no longer), MM thought
it is possible that the ``Al$_{11}$Ir$_4$'' structure reported by 
Tamura~\cite{tamura}
to have a $2\times 2 \times 2$ supercell and $Pa\bar{3}$ space group
was actually our Al$_{41}$Ir$_{23}$ phase, due to loss of Al content
in the sample preparation.}

\SAVE{MM 12/3/11 says: most likely, the Al11Ir4 should transform immediately
below the structural ordering transition temperature, 
at which it loses most of the entropy.}

\secc{Electronic density of states}
A specific interest in Al--Ir and certain other Al--TM systems is
the possibility of an insulating alloy, the elemental constituents
of which are all good metals,   e.g. Al$_2$Ru in the
TiSi$_2$ structure~\cite{Al2Ru}
Al$_2$Fe in the MoSi$_2$ type-structure~\cite{Al2Ru,MM-MW-Al2Fe}.
Quasicrystal $i$-AlPdRe (built from pMI clusters)
is long claimed to be a semiconductor~\cite{poon-AlPdRe,krajci-hafner-AlPdRe}.
Predicting gap formation in these alloys
depends critically on the accurate relaxation of atomic positions
~\cite{krajci-hafner-AlPdRe}.
\LATER{Marek: in the original draft you cited "Krajci and Hafner" for this
last citation; I made a dummy reference, but you never told me which paper 
you had in mind.}

\begin{figure}
\vskip 1.0cm 
\includegraphics[width=3.2in]{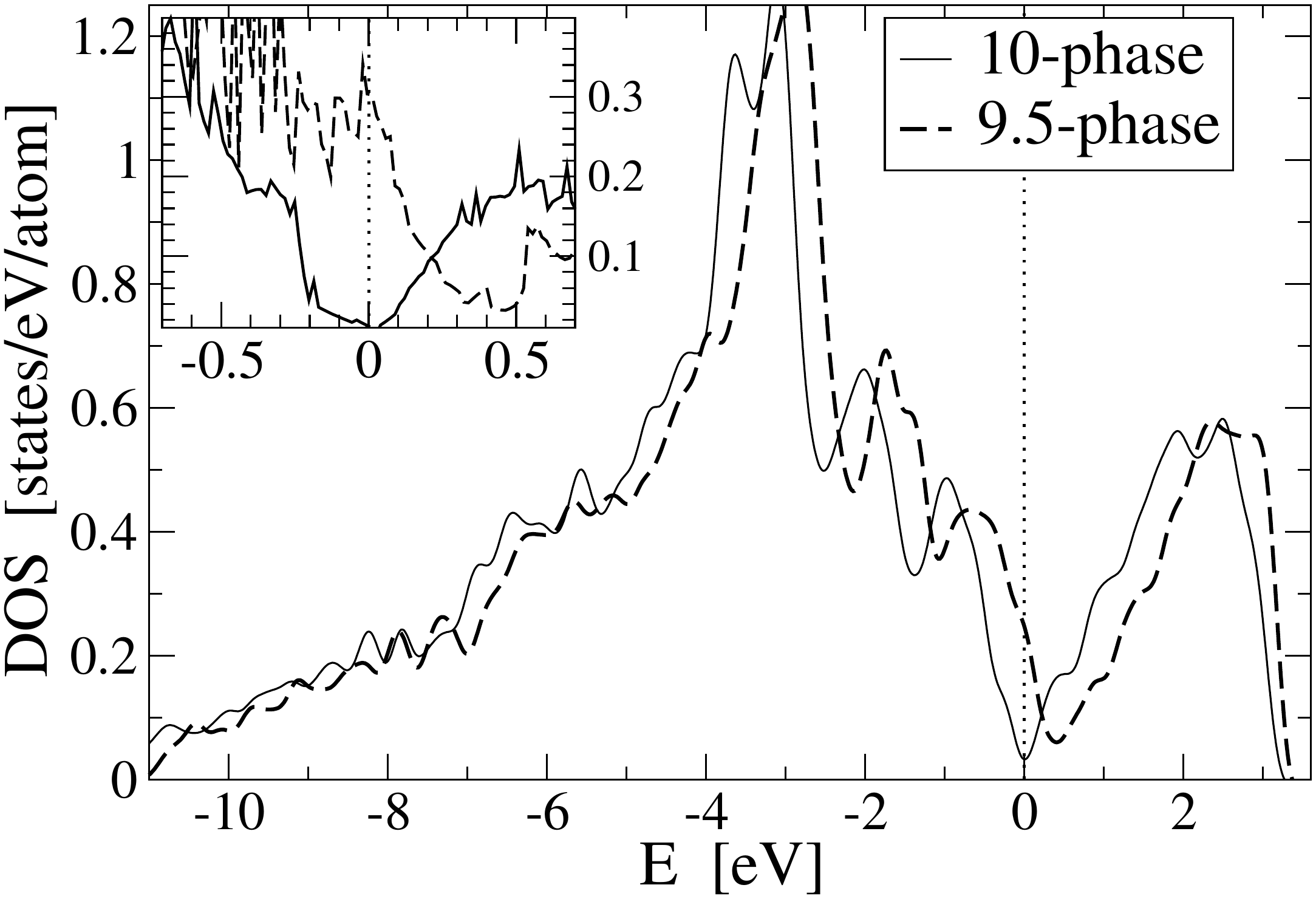}
\caption{ Electronic density of states (DOS) of $T=0$ ordered
  structures, where $E=0$ is the Fermi energy; (curves are smoothed
  using the Methfessel-Paxton smearing method).  \SAVE{The ``10''
    phase (solid line) and ``9.5'' phase (dashed).}  The 10-phase has
  a gap of $\sim 40$ meV (blown up in inset, tetrahedron method) but
  in the 9.5 phase, the Fermi level is $\sim 0.5$ eV shifted from a
  pseudogap.
}
\label{fig:edos}
\end{figure}

We find the Al$_{11}$Ir$_4$ electronic density of states (DOS) 
in the ordered ($T=0$) ``10'' phase has a narrow gap, and significant pseudogap
as shown in Fig.~\ref{fig:edos}, so this alloy should be a semiconductor. 
By contrast, the ordered ``9.5'' phase is predicted to be robustly metallic
[Fig.~\ref{fig:edos}(b)] 
(In the disordered high-temperature ``10'' phase [not shown]
the gap tends to get filled in, so that phase is also likely
to be metallic.)
\SAVE{But all in all, the electronic DOS of the high-$T$ 10-phase is very similar to
the low-$T$ 10-phase,  as compared to
the big difference between the 9.5 and 10 phases.}
We call attention to the fact that 
these very similar structures have radically different electronic properties,
all due to details of the placement of certain Al atoms that link 
adjacent clusters.

\secc{Discussion}
In conclusion, starting from diffraction--data refined
average structure of Al$_{11}$Ir$_4$\cite{grin}, we combined 
molecular dynamics simulations with pair potentials
to discover a well-defined, asymmetric inner 
Al$_{9}$Ir and Al$_{10}$Ir clusters,
with variable orientations, and an ordering transition to new orthorhombic phases.
We suggest an experimental search for the ordered phase having all clusters
of Al$_{10}$Ir type, as it has a {\it gap} exactly at the Fermi energy, and hence 
should be a semiconductor.
We also found that, contrary to the 
accepted phase diagram, the stable low-temperature phase at $x_{\rm Ir}\approx 1/4$ is
Al$_{21}$Ir$_8$.

Several other known structures are made by placing
pMI clusters and/or variants on the
same simple cubic lattice with $a\sim$7.7\AA:
not only Al$_{41}$Cu$_8$Ir$_{15}$,~\cite{AlIrCu},
Al$_{68}$Pd$_{20}$Ru$_{12}$~\cite{mahne-steurer} 
and AlCuRuSi~\cite{sugiyama-AlCuRuSi}, which we
mentioned, but also
Al$_{70}$Pd$_{10}$Fe$_{20}$~\cite{grushko-AlPdFe},
\QOK{Now I am wondering if the AlPdFe structure 
mentioned is actually the same pMI's as in the
AlCuIr structure, or is made of other kinds of
cluster centers, as in all the other cubic phases
we mention.  Please check!}
\SAVE{The compounds listed are
all (pseudo)ternaries from the AlCuFe/AlPdMn family.
They are known from the
quasicrystal viewpoint~\cite{elser-henley} as ``1/0 approximants''.}
and $\epsilon$--Ag$_7$Mg$_{26}$ (in the latter case, 
the ``even''-node inner clusters are AgMg$_8$ cube, 
while ``odd'' nodes are perfect Mackay icosahedra).
\SAVE{(``The odd, i.e.  body-centered clusters,
are perfect Mackay icosahedra in this case;  in all
the other structures they are complete MI, but
usually their outer shells are more distorted.}
\SAVE{(These MI have a filled center, so the inner cluster is AgMg$_{12}$)}
\SAVE{Notice this cube is OPPOSITE to the AlCu$_8$ cube variant,
since Mg is analogous to Al (CLH: but both examples are fragments
of the B2 structure).
The central Ag has 14 Mg neighbors in the
form of a bcc coordination shell.
First, eight Mg in the cube are at 2.9\AA from the center.
They are followed by six Mg along 2--fold
axes at 3.8\AA from the center, which are
part of the linkages to the adjacent even clusters, and 
thus dimpled.  These 2--fold axis atoms have to be large atom, 
since they are shared with second shell of the neighboring MI.}

Furthermore, the lowest energy version yet found of
the $i$-AlPdMn quasicrystal structure~\cite{krajci-AlPdMn,MM-AlPdMn}
consists of exactly the same pMI clusters described above, in particular (i)
they have a short ($\sim 8$\AA) linkage along the icosahedral twofold
direction, rather than the $\sim 12.5$\AA~ linkage known from
$\alpha$-AlMnSi~\cite{elser-henley} (ii) they have 
an Al$_{10}$Mn inner cluster in the same shape as the
one described here, which appear in various orientations
with only small energy differences.
The decagonal-related $\xi'$(AlPdMn) phase contains pMI
clusters with Al$_9$Mn and Al$_{10}$Mn inner clusters,
not quite identical to those we presented here.
\SAVE{In the decagonal type phases,
the Al$_9$Mn inner cluster is trigonal, like our Al$_{10}$Ir with the end
cap removed; the Al$_{10}$Mn is square, like our Al$_{9}$Ir
but capped at {\it both} ends.}

\SAVE{(comment from 12/3/11).
``The existence of the Al$_{21}$Ir$_8$ phase, built from Al$_{10}$Ir 
inner clusters, shows that this cluster is a stable motif that does not
depend on the second icosahedral shell (and strikingly, after shared
atoms are taken into account, Al$_{21}$Ir$_8$ has practically the same 
composition as Al$_{11}$Ir$_4$).'' MM : OK.}

The icosahedral cage containing an inner cluster that
breaks icosahedral symmetry is known in a quite
different family of quasicrystals and related compounds:
CaCd$_6$ or ScZn$_6$, with an inner 
Cd$_4$ or Zn$_4$ tetrahedron~\cite{gomez,MM-CLH-Zn4}.
Those, too, show orientational orderings at low temperatures~\cite{CaCd-order,ishii-CaCd-order}.
We observe that, if there is any way at all in icosahedral quasicrystals
to implement the long-sought ``local matching rules'' that would 
stabilize an ideal quasiperiodic ground state, or merely rules that
limit tiling randomness to large spatial scales 
\SAVE{(by the constraint of a supertiling)}
the likeliest candidate are these inner clusters that
spoil the clusters' high symmetry.
\SAVE{(much as the Penrose arrows spoil the equivalence of
different rhombus edges).}

On the theoretical side, this work on Al--Ir could be extended in the following directions:
(1) simulation of the inner-cluster dynamics for comparison
to neutron data,
\SAVE{of the dynamic structure factor,}
analogous to the Cd$_4$ tetrahedron in
CaCd$_6$ ~\cite{MM-CLH-Zn4};
(2) fitting an orientational interaction between
adjacent clusters~\cite{justin-thesis}
\SAVE{(as was done for the Cd$_4$--Cd$_4$ interaction~\cite{brommer}),}
for Monte Carlo estimation
of the orientational $T_c$.
\SAVE{Anything would be more accurate than 
the $2\times 2\times 2$ supercell used here.}
(3) a possible Al--Ir quasicrystal 
\SAVE{(or related large unit cell phases)}
based on these clusters. 
Experimental clarification of the low--temperature
stabilities of Al--Ir around $x_{Ir}\sim 0.27$ are of course highly desirable.

\SAVE{An effective Hamiltonian describing the phase transition would consist of
cluster--cluster interactions. Taking into account cluster symmetry,
we find there are 28 independent relative orientations of two adjacent
interior  clusters.  This is left for a later work.}

\acknowledgments We thank W. Steurer and P. Kuczera for sharing
structural data on Al$_{41}$Cu$_8$Ir$_{15}$ phase prior to
publication, R. Tamura for fruitful discussions, and Yu. Grin for
careful reading of the manuscript and comments, in particular pointing
out cluster relationship with Al$_6$Mn structure.  This work was
supported by DOE Grant DE-FG02-89ER-45405 (MM, CLH), and Slovak
funding VEGA 2/0111/11 and APVV-0492-11 (MM).

%
%
%
%
%
%
%
%
%

\section{Supplementary information}

In this section, we give tables of three kinds of detailed
information omitted from the main text of the paper.

\subsection*{Al-Ir phases for stability calculations}

Table \ref{tab:pdiag} gives the data used to find 
the predicted binary phase diagram of the Al--Ir. system,
by the usual convex-hull construction.
Despite the structure names, a binary Al--Ir composition
is used in all cases, with Ir replacing any transition metals.
All energies are ab--initio calculations with the VASP code.

\begin{table}[b]
\begin{tabular}{ccccc}
  structure &  $\Delta E$ & $\Delta H$ &  x(Al) & x(Ir) \\
 & meV/at.&meV/at.& \% & \%\\
\hline
  Al$_9$Co$_2$.mP22        &   0 & -544.0 &   81.8 &   18.2 \\   
  Al$_{45}$Ir$_{13}$.oP236  &  0 & -658.8 &   77.6 &   22.4 \\  
  Al$_{21}$Pd$_8$.tI116    &  0 & -759.6 &   72.4 &   27.6 \\  
  AlIr.cP2                 & 0 & -955.3 &   50.0 &   50.0 \\  
\hline
 Al$_5$Co$_2$.hP28         &    2.8 & -765.4 &   71.4 &   28.6 \\
 Al$_{11}$Ir$_4$.oP60\footnotemark[1]   &    3.7 & -737.9 &   73.3 &   26.7 \\
 Al$_{41}$Ir$_{23}$.hR64    &    5.6 & -826.9 &   64.1 &   35.9 \\
 Al$_{28}$Ir$_9$.hP236     &  7.2 & -672.3 &   76.5 &   23.5 \\    
 Al$_3$Ir.hP24\footnotemark[2]   &    7.8 & -722.5 &   73.9 &   26.1 \\
 Al$_{11}$Ir$_4$.oA236\footnotemark[3]  &    9.2 & -741.2 &   72.9 &   27.1 \\
 Al$_3$Ir.hP8              &   25.0 & -684.2 &   75.0 &   25.0 \\
\hline
\end{tabular}\\
\footnotetext[1] {~ ``10-phase''}
\footnotetext[2]{~ triple supercell of Al$_3$Ir, with one less Al atom (site vacant).}
\footnotetext[3] {~ ``9.5-phase''}
\caption{ \label{tab:pdiag}
Energy diagram of Al--Ir system at T=0K. Column $\Delta E$ 
is energy in meV/atom by which a structure is unstable relative to a mixture of
competing stable compounds. Stable compounds forming convex hull of energy--composition
scatter plot have $\Delta E=0$.
}
\end{table}

\subsection*{Fitted pair potentials}

\begin{figure}[b]
\vskip 1.0cm   
\includegraphics[width=3.2in]{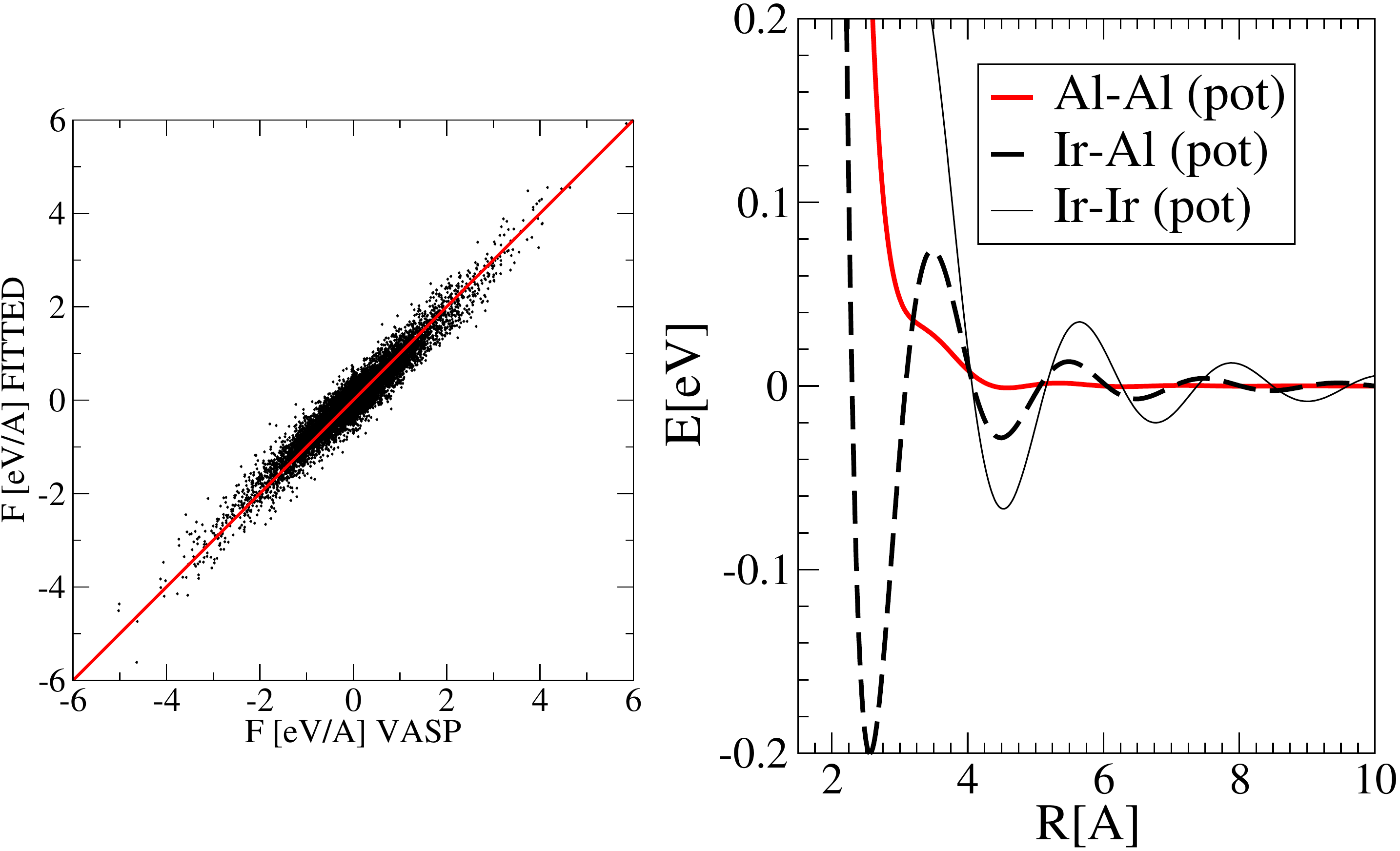}
\caption{ 
\label{fig:ppfit}
EOPP potentials fitted to ab--initio (VASP) force and
energy datapoints. Fit to forces (left panel) is shown with fitted
forces (vertical axis) plotted against VASP forces (horizontal axis). 
The fitted pair potentials are shown in the right panel. Parameters
}
\end{figure}

Most calculations in the  paper were based on em
``empirical oscillating pair potentials''  (EOPP) of form
\begin{equation}
  \label{eq:oscil6}
    V(r) = \frac{C_1}{r^{\eta_1}} + \frac{C_2} {r^{\eta_2}} \cos(k_* r + \phi_*)
\end{equation}
as presented by Mihalkovi\v{c} and Henley (main text Ref. [6]).
where $r$ is the distance between a pair of atoms. 
The fitted coefficients we used are listed in Table \ref{tab:eopp}.
Fig. \ref{fig:ppfit} gives a scatter plot demonstrating the
goodness of fit and a plot of the potentials themselves.
As is typical of Al-transition metal (Al-TM) potentials,
with TM=Ir in this case:
(a) the Al-Al potential has no nearest-neighbor well but only
a shoulder (b) the Al-Ir potential has a very deep well
at the nearest-neighbor distance (c) the Ir-Ir potential
is unfavorable for nearest neighbors but has a deep well
at the second neighbor distance (d) all the potentials
have relatively strong Friedel oscillations.

\begin{table}[t]
\begin{tabular}{c|cccccc}
\hline
 & $C_1$ & $\eta_1$ & $C_2$ & $\eta_2$ & $k_*$ & $\Phi_*$ \\ 
\hline
Al--Al &     9.9811$\times$ 10$^2$&  8.7044&  -11.3044&   5.4518&  3.5646 & 2.5119 \\
Al--Ir &   1.3764$\times$ 10$^4$& 13.3271&    9.2649&  3.8320&  3.1919 & 1.1091 \\
Ir--Ir &  3.2083$\times$ 10$^5$& 12.7731&   -7.2818&  3.0765& -2.8274 & 0.4493 \\
\end{tabular}
\caption{\label{tab:eopp} Fitted parameters for Al--Ir EOPP potentials.}
\end{table}

\section*{Crystal structure of phase with Al$_{10}$Ir clusters}

Table \ref{tab:10phase}. lists all the
Wyckoff positions of the low--temperature
structure of the ``10-phase'' of Al$_{11}$Ir$_4$;
The cluster orientations correspond to Figure 1(b) 
of the main text, corresponding to a cubic cell doubled in
one direction (becoming the $a$ direction); since the
cubic structure lacks 4-fold symmetry, this causes the
$b$ and $c$ cell parameters to become different by
$\sim 0.6$\%.

\begin{table}
\small
\begin{tabular}{cccc|cccc}
site & X  & Y  & Z  &
site & X  & Y  & Z  \\ 
\hline
   Ir1/2b &    0.5000 &    0.5000  &   0.5135&
   Ir2/2c &    0.2500 &    0.9952  &   0.0140\\
   Ir3/2c &    0.2500 &    0.4745  &   0.2897&
   Ir4/2c &    0.2500 &    0.5099  &   0.7156\\
   Ir5/4d &    0.5028 &    0.7269  &   0.0113&
   Ir6/4d &    0.3868 &    0.9996  &   0.5274\\
   Al1/4d &    0.5006 &    0.7884  &   0.6822&
   Al2/4d &    0.3544 &    0.3333  &   0.5161\\
   Al3/4d &    0.3521 &    0.6752  &   0.5022&
   Al4/4d &    0.5850 &    0.4755  &   0.2149\\
   Al5/4d &    0.5898 &    0.4900  &   0.8229&
   Al6/4d &    0.5175 &    0.8086  &   0.3328\\
   Al7/4d &    0.3668 &    0.8794  &   0.2035&
   Al8/2c &    0.2500 &    0.1771  &   0.7409\\
   Al9/4d &    0.4059 &    0.9910  &   0.8849&
  Al10/2c &    0.7500 &    0.1536  &   0.7156\\
  Al11/4d &    0.6644 &    0.7314  &   0.0690&
  Al12/2c &    0.2500 &    0.0944  &   0.3570\\
  Al13/2c &    0.2500 &    0.6492  &   0.0134& &&&\\
\hline
\end{tabular}\\
\caption{\label{tab:10phase} List of Wyckoff sites for ``10-phase'' structure.
Space group is $Pma2$ (\#28), Pearson symbol $oP60$. 
VASP-optimized lattice parameters are $a$=15.492\AA, $b$=7.725\AA and
$c$=7.685\AA.
}
\end{table}

\begin{table}
\small
\begin{tabular}{cccc|cccc}
site & X  & Y  & Z  &
site & X  & Y  & Z  \\ 
\hline
    Ir1/8d &    0.7498 &   0.0006 &   0.0008 & 
    Ir2/4c &    0.0003 &   0.2500 &   0.7498 \\ 
    Ir3/4c &    0.7499 &   0.2500 &   0.6054 & 
    Ir4/8d &    0.3602 &   0.0013 &   0.7517 \\ 
    Ir5/8d &    0.5001 &   0.8857 &   0.4935 & 
    Ir6/4c &    0.7506 &   0.7500 &   0.4015 \\ 
    Ir7/4c &    0.5018 &   0.7500 &   0.2503 & 
    Ir8/4c &    0.2619 &   0.2500 &   0.6151 \\ 
    Ir9/4c &    0.2461 &   0.7500 &   0.3962 & 
   Ir10/8d &    0.8581 &   0.9789 &   0.2506 \\ 
   Ir11/8d &    0.0006 &   0.3900 &   0.4985 & 
    Al1/8d &    0.3968 &   0.0026 &   0.9159 \\ 
    Al2/8d &    0.1609 &   0.3489 &   0.5097 & 
    Al3/8d &    0.2589 &   0.0888 &   0.1491 \\ 
    Al4/8d &    0.2355 &   0.9163 &   0.1534 & 
    Al5/8d &    0.1613 &   0.6470 &   0.4972 \\ 
    Al6/8d &    0.2433 &   0.0876 &   0.8458 & 
    Al7/8d &    0.3353 &   0.6469 &   0.5003 \\ 
    Al8/8d &    0.3340 &   0.3545 &   0.5049 & 
    Al9/8d &    0.2591 &   0.9076 &   0.8499 \\ 
   Al10/8d &    0.4067 &   0.9853 &   0.0921 & 
   Al11/8d &    0.9012 &   0.8370 &   0.3482 \\ 
   Al12/4c &    0.5471 &   0.7500 &   0.5829 & 
   Al13/8d &    0.4266 &   0.8563 &   0.6523 \\ 
   Al14/4c &    0.4199 &   0.2500 &   0.3969 & 
   Al15/4c &    0.5860 &   0.2500 &   0.3900 \\ 
   Al16/8d &    0.4944 &   0.9075 &   0.8078 & 
   Al17/4c &    0.3149 &   0.2500 &   0.2534 \\ 
   Al18/8d &    0.9035 &   0.9986 &   0.9101 & 
   Al19/8d &    0.9061 &   0.9990 &   0.0891 \\ 
   Al20/8d &    0.3622 &   0.8352 &   0.2277 & 
   Al21/8d &    0.0296 &   0.9100 &   0.2485 \\ 
   Al22/4c &    0.0351 &   0.2500 &   0.5873 & 
   Al23/8d &    0.8927 &   0.1643 &   0.6627 \\ 
   Al24/4c &    0.1662 &   0.7500 &   0.2510 & 
   Al25/4c &    0.0534 &   0.7500 &   0.4062 \\ 
\hline
\end{tabular}\\
\caption{\label{tab:oa236} Atomic structure of ``9.5-phase'' ($oA236$)
in space group $Abm2$ (\# 39). Optimized lattice parameters are
$a$=15.321~\AA, $b$=15.475~\AA, $c$=15.412~\AA. Cluster centers
are Ir2 (``9-clusters'') and Ir7 (``10-clusters'').
}
\end{table}

\begin{table}
\small
\begin{tabular}{cccc|cccc}
site & X  & Y  & Z  &
site & X  & Y  & Z  \\ 
\hline
    Ir1/4b &    0.5000 &   0.0000 &   0.2500 & 
    Ir2/8j &    0.2497 &   0.2507 &   0.5000 \\ 
    Ir3/8j &    0.9954 &   0.3955 &   0.5000 & 
   Ir4/16k &    0.3926 &   0.2502 &   0.7704 \\ 
   Ir5/16k &    0.2503 &   0.5011 &   0.3603 & 
    Ir6/8j &    0.9997 &   0.1036 &   0.5000 \\ 
    Ir7/4a &    0.0000 &   0.0000 &   0.7500 & 
   Al1/16k &    0.8442 &   0.5900 &   0.2517 \\ 
   Al2/16k &    0.0877 &   0.9954 &   0.3987 & 
   Al3/16k &    0.4831 &   0.8460 &   0.1605 \\ 
   Al4/16k &    0.9124 &   0.5012 &   0.1026 & 
   Al5/16k &    0.9870 &   0.1541 &   0.3407 \\ 
   Al6/16k &    0.1544 &   0.0893 &   0.7501 & 
   Al7/16k &    0.1490 &   0.1544 &   0.4135 \\ 
   Al8/16k &    0.1403 &   0.3379 &   0.4147 & 
   Al9/16k &    0.2780 &   0.2536 &   0.6603 \\ 
   Al10/8j &    0.2925 &   0.4107 &   0.5000 & 
   Al11/8j &    0.3035 &   0.0941 &   0.5000 \\ 
   Al12/8j &    0.4193 &   0.2521 &   0.5000 & &&& \\ 
\hline
\end{tabular}\\
\caption{\label{tab:oi232} ``9-phase'' structure ($oI232$),
space group $Ibam$ (\# 72),
orthorhombic lattice parameters are $a$=15.136~\AA,
$b$=15.517~\AA, $c$=15.457~\AA. ``9-cluster'' centers are
at Ir2 site.}
\end{table}

\begin{table}
\small
\begin{tabular}{cccc|cccc}
site & X  & Y  & Z  &
site & X  & Y  & Z  \\ 
\hline
    Ir1/3a &    0.0000 &   0.0000 &   0.0020 & 
    Ir2/3a &    0.0000 &   0.0000 &   0.7488 \\ 
    Ir3/3a &    0.0000 &   0.0000 &   0.2496 & 
    Ir4/9b &    0.2685 &   0.2430 &   0.6356 \\ 
    Ir5/9b &    0.6312 &   0.0770 &   0.4618 & 
    Ir6/9b &    0.2696 &   0.9981 &   0.3618 \\ 
    Ir7/9b &    0.3639 &   0.9178 &   0.5390 & 
    Ir8/3a &    0.0000 &   0.0000 &   0.5959 \\ 
    Ir9/9b &    0.7434 &   0.8712 &   0.4665 & 
   Ir10/3a &    0.0000 &   0.0000 &   0.4027 \\ 
   Ir11/9b &    0.2580 &   0.1292 &   0.5313 & 
    Al1/3a &    0.0000 &   0.0000 &   0.4995 \\ 
    Al2/9b &    0.8717 &   0.1251 &   0.4363 & 
    Al3/9b &    0.1273 &   0.2493 &   0.5613 \\ 
    Al4/9b &    0.2697 &   0.2245 &   0.2304 & 
    Al5/9b &    0.0399 &   0.2553 &   0.2678 \\ 
    Al6/9b &    0.0192 &   0.1585 &   0.1729 & 
    Al7/9b &    0.1447 &   0.1397 &   0.3285 \\ 
    Al8/9b &    0.7519 &   0.7866 &   0.7716 & 
    Al9/9b &    0.9571 &   0.7422 &   0.7301 \\ 
   Al10/9b &    0.9820 &   0.8361 &   0.8292 & 
   Al11/9b &    0.8418 &   0.8617 &   0.6704 \\ 
   Al12/9b &    0.5229 &   0.1744 &   0.4054 & 
   Al13/9b &    0.1543 &   0.4314 &   0.6314 \\ 
   Al14/9b &    0.8615 &   0.5862 &   0.3442 & 
   Al15/3a &    0.0000 &   0.0000 &   0.9080 \\ 
\hline
\end{tabular}\\
\caption{\label{tab:hr64} Structure of rhombohedral Al$_{41}$Ir$_{23}$ ($hR64$) phase,
space group $R3$ (\# 146) in hexagonal setting.
Optimized lattice parameters are $a$=11.032~\AA and $c$=27.073~\AA.
``10-clusters'' surround Ir1 sites, Al1 site is center of``B2''-structure local environments.
}
\end{table}



\begin{thebibliography}{99}
\bibitem{FN-rc}
The portion of the potentials
between 7--10\AA~ is crucial for (even mechanical) stability.
\SAVE{when $r_c=7$\AA~ was used, an entirely different low-T structure 
was obtained, with the backbone of large Ir icosahedra highly distorted.
MM forgets the details.}
\bibitem{FN-supercells}
Supercells are essential here, as adjacent clusters have
different orientations in the low-energy states.
\bibitem{FN-Grin-spacegroups}
{Of the two possible space groups suggested in Ref.~\cite{grin},
we must use $Pm\bar{3}$, which has inversion symmetry.}
\bibitem{Al6Mn}
Y. Grin has pointed out that our 10-cluster is identical
to the uncaged cluster from which Al$_6$Mn is built
[A. D. I. Nickol, Acta Crystallogr. 6, 285 (1953)].
\end{thebibliography}
\end{document}